\documentclass[draft,jgrga]{agutex}

 \usepackage{lineno}
 \linenumbers*[1]

  \usepackage[dvips]{graphicx}
\authorrunninghead{HIGASHIMORI AND HOSHINO}

\titlerunninghead{SLOW SHOCKS AND ION TEMPERATURE ANISOTROPY}

\authoraddr{K. Higashimori,
Department of Earth and Planetary Science, University of Tokyo, Science
Building 1, Hongo, 7-3-1, Japan.
(higashi@eps.s.u-tokyo.ac.jp)
}

\authoraddr{M. Hoshino,
Department of Earth and Planetary Science, University of Tokyo, Science
Building 1, Hongo, 7-3-1, Japan.
(hoshino@eps.s.u-tokyo.ac.jp)
}

\begin{document}

%
%

\title{The Relation between Ion Temperature Anisotropy and Formation of
Slow Shocks in Collisionless Magnetic Reconnection}
%

%
%



\authors{
K. Higashimori\altaffilmark{1} and M. Hoshino\altaffilmark{1}
}

\altaffiltext{1}{
Department of Earth and Planetary Science, University of Tokyo, Tokyo, Japan.
}

%
%


\begin{abstract}
 We perform a two-dimensional simulation by using an electromagnetic
 hybrid code to study the formation of slow-mode shocks in collisionless
 magnetic reconnection in low beta plasmas, and we focus on the
 relation between the formation of slow shocks and the ion temperature
 anisotropy enhanced at the shock downstream region.
 It is known that as magnetic
 reconnection develops, the parallel
 temperature along the magnetic field becomes large in association with
 the anisotropic PSBL
 (plasma sheet boundary layer)
 ion beams, and this temperature anisotropy has a
 tendency to suppress the formation of slow shocks.
 Based on our simulation result, we found that
 the slow shock formation is suppressed
 due to the large temperature anisotropy near the X-type
 region, but the ion temperature anisotropy relaxes with
 increasing the distance from the magnetic neutral point.
 As a result, two pairs of current structures, which are the strong
 evidence of dissipation of magnetic field in slow shocks, are formed at
 the distance $\left|x\right|\geq 115\ \lambda_{\rm i}$ from the neutral
 point.
\end{abstract}

%
%

%

\begin{article}

%
%

 \section{Introduction}

 In the Earth's magnetotail, magnetic reconnection plays an important
 role in the conversion of magnetic field energy in two lobes into
 kinetic and thermal energy of plasmas in a plasma sheet.
 Since \cite{Petschek1964} proposed the necessity of pairs of slow-mode
 shocks attached to the diffusion region in magnetic reconnection in
 order to achieve its efficient energy conversion rate, many studies have
 been devoted to this model.
 As for numerical studies, many MHD simulations have confirmed the
 existence of slow shocks along the reconnection layer.
 Namely, slow-mode waves, propagating from the
 neutral point toward two lobes and along the outflow jets at one time,
 steepen and result in the steady state two pairs of slow shocks
 as is suggested by Petscheck \citep{Sato1979, Scholer1989, Abe2001}.
 In addition, observations of ISEE
 \citep{Feldman1985} and Geotail \citep{Saito1995, Seon1995} satellites
 have shown the existence of such slow shocks in the Earth's magnetotail.
 Especially, \cite{Saito1995} showed that the variation in the
 ion temperature is much larger than the electron temperature variation
 across slow shocks.
 This suggests that the ion scale dissipation mechanism would be strongly
 related to the formation of slow shocks in collisionless plasmas.

 Until now, however, there is no clear consensus on the formation of
 slow shocks
 in magnetic reconnection by both hybrid and full-particle simulations,
 even though the formation of slow shocks itself has been demonstrated by
 slow shock simulations without magnetic reconnection
 \citep{Omidi1989, Karimabadi1995B, Omidi1995}, and by
 a Riemann problem
 of slow shocks \citep{Fujimoto1994,Lin1995,Liu2011}.
 Attempts to investigate slow shocks in a large scale reconnection
 with two-dimensional hybrid codes were first done by \cite{Krauss1995}.
 They showed a fine structure of the reconnection layer where
 the fast plasma flow is generated.
 They concluded that
 these transition layers did not confirm the properties of the expected
 slow shocks, and mentioned that the reason for this is due to the fact
 that the ion dissipation scale is comparable to the thickness of the
 developing plasma sheet.
 \cite{Lin1996} and \cite{Lottermoser1998} also performed large
 scale hybrid simulations to investigate slow shocks in magnetic reconnection.
 \cite{Lin1996} suggested that the isotropic Rankine-Hugoniot (RH) jump
 conditions of slow shocks were better satisfied
 with increasing distance
 from the neutral point in the case of no-guide field reconnection.
 \cite{Lottermoser1998} also showed the slow shock-like discontinuities,
 and suggested that the downstream ions were not directly
 heated by slow shocks as was shown by \cite{Lin1996} but heated by the
 stochastic motion of ions.
 They discussed that the thin current sheet formed after
 reconnection became turbulent and such turbulent structures
 caused stochastic motion of ions.

 That being the case, what causes such a discrepancy of the
 formation of slow shocks between MHD \citep{Sato1979} and kinetic
 treatments of plasmas?
 We suggest that an important kinetic modification in magnetic
 reconnection would be ion temperature anisotropy along reconnection layers.
 Observations by ISEE-3 and Geotail satellites in the Earth's
 magnetotail reported that the ion temperature
 parallel to the magnetic field ($T_{\rm i,\parallel}$) is higher than
 that perpendicular to the magnetic field ($T_{\rm i,\perp}$) at the
 downstream region of slow shocks \citep{Cowley1984,Hoshino2000}.
 Such ion temperature anisotropy is known to be produced by two
 plasma components, i.e., the convecting cold lobe ions with the velocity
 ${\bf V}_d=c{\bf E}\times{\bf B}/B^2$ and the PSBL (plasma sheet
 boundary layer) ion beams which are accelerated and ejected from the
 diffusion region.
 Preceding studies for magnetic reconnection by kinetic simulations also
 support the fact that $T_{\rm i,\parallel}/T_{\rm i,\perp}>1$ at the
 downstream of discontinuities along the reconnection layer
 \citep{Hoshino1998,Lottermoser1998}.
 On the other hand, theoretical studies about slow shocks with the
 one-dimensional RH relations suggest that
 conditions required to satisfy slow shock RH relations become
 restricted by the ion temperature anisotropy,
 $T_{\rm i,\parallel}/T_{\rm i,\perp}>1$, in
 the downstream region
 \citep{Lyu1986, Karimabadi1995A}.
 They showed that slow shock solutions in anisotropic plasmas are
 greatly affected by the downstream temperature anisotropy.
 If $T_{\rm i,\parallel}/T_{\rm i,\perp}>1$ in the downstream region, 
 slow shock solutions can exist only for the limited upstream Mach
 number regime.
 This means that the enhancement of ion temperature anisotropy along the
 reconnection layer makes it harder for slow shocks to exist in magnetic
 reconnection.
 In this study, we focus on this point and perform kinetic simulations
 for magnetic reconnection.
 Then, the relation between such ion temperature
 anisotropy and formation of slow shocks in magnetic reconnection is
 discussed in accordance with anisotropic RH relations.

 In the following sections, we will first refer to the simulation model;
 second, show results of the kinetic simulation for magnetic
 reconnection, and lastly discuss the nature of discontinuities formed
 along reconnection layers in detail.

 \section{Simulation Model}
 
 In our study, the two-dimensional electromagnetic hybrid code, in which
 ions are treated as particles while electrons as a mass-less fluid, is
 used to investigate the formation of slow shocks in magnetic reconnection.
 Algorithm of our hybrid code is based on the method of general
 predictor-corrector loops \citep{Harned1982, Winske1984}.
 Within our simulation, basic equations are as follows:
 \begin{eqnarray}
  m_{\rm i}\frac{d{\bf v}_{\rm i}}{dt} = q_{\rm i}
   \left( {\bf E} + \frac{{\bf v}_{\rm i}\times{\bf B}}{c} \right)
   , & \\
  \frac{\partial {\bf B}}{\partial t} = -c\nabla\times{\bf E}
  , & \\
  {\bf E} = -\frac{1}{q_{\rm i} n_{\rm i}}\nabla p_{\rm e} -
  \frac{1}{c}{\bf V}_{\rm e}\times{\bf B} + \eta{\bf J}
  , & \\
  {\bf V}_{\rm e} = {\bf V}_{\rm i} - \frac{c}{4\pi en_{\rm
  i}}\nabla\times{\bf B}
  , & \\
  {\bf V}_{\rm i} = \frac{\displaystyle
  \int_{-\infty}^{\infty}{\bf v}_{\rm i} f_{\rm i}({\bf v}_{\rm
  i})d{\bf v}_{\rm i}}{\displaystyle
  \int_{-\infty}^{\infty}f_{\rm i}({\bf v}_{\rm i})d{\bf v}_{\rm i}}
  , & \\
  \frac{\partial p_{\rm e}}{\partial t} + 
  \left({\bf V}_{\rm e}\cdot\nabla\right) p_{\rm e}
  + \gamma p_{\rm e}\left(\nabla\cdot{\bf V}_{\rm e}\right)
  - \left(\gamma-1\right)\eta{\bf J}^2 = 0
  ,
 \end{eqnarray}
 where $f_{\rm i}$ is the velocity distribution function of ions and a
 charge neutral condition, $q_{\rm i} n_{\rm i} - en_{\rm e} = 0$, is
 assumed.
 $\gamma$ is the adiabatic index and set to be $5/3$ in this paper.
 As for electrons, we assume that the electron gas is isotropic and both
 the electron heat flux and the viscosity stress tensor are neglected.
 In our simulation, various parameters are normalized by the
 parameters in the initial lobe (e.g., $n=n/n_0$, ${\bf B}={\bf B}/B_0$,
 and ${\bf V}={\bf V}/V_{A0}$, where $n_0$, $B_0$, and $V_{A0}$ are
 respectively the density, the magnetic field, and the
 Alfv$\acute{\rm e}$n velocity in the initial lobe).

 In addition, the spatial profile of the resistivity is given by
 \begin{equation}
  \eta\left(x,y\right) = \eta_0 +
   \eta_c\cosh^{-2}\left[
		    \left(\frac{x}{l_x}\right)^2+\left(\frac{y}{l_y}\right)^2
		   \right],
 \end{equation}
 where $\eta_0$ is the background resistivity due to ion-electron
 interactions.
 Here, the background resistive length
 $\lambda_{{\rm r}0}\equiv\eta_0c^2/(4\pi V_{A0})$ is set to be
 $10^{-4}\ \lambda_{\rm i}$, where $\lambda_{\rm i}$ is the ion inertial
 length in the initial lobe.
 $\eta_c$ is the anomalous resistivity due to some instabilities in
 the diffusion region, e.g., the lower hybrid drift and the drift kink
 instabilities.
 This resistive term is assumed to be independent of time and its
 resistive length is equal to $3.6\times 10^{-2}\ \lambda_{\rm i}$.
 $l_x$ and $l_y$ are the characteristic lengths to determine the size of
 the anomalous (electron) diffusion region and are set to be
 $l_x=1.0\ \lambda_{\rm i}$ and $l_y=0.5\ \lambda_{\rm i}$, respectively.

 The whole size of the two-dimensional system is
 $-342\ \lambda_{\rm i}\leq L_x\leq 342\ \lambda_{\rm i}$ and
 $-32\ \lambda_{\rm i}\leq L_y\leq 96\ \lambda_{\rm i}$.
 Grid intervals, $\Delta_x$ and $\Delta_y$, are both equal to
 $1/3\ \lambda_{\rm i}$.
 Initially, a double Harris equilibrium is assumed 
 and periodic boundary conditions in both $x$- and $y$-direction
 are imposed.
 The initial spatial profile of magnetic field is given as
 \begin{equation}
  {\bf B}(y) = B_0\left[\tanh\left(\frac{y}{\delta_y}\right)
		   -\tanh\left(\frac{y-y_c'}{\delta_y}\right)-1\right]{\bf e}_x,
 \end{equation}
 where $y_c'=64.0\ \lambda_{\rm i}$ and the half thickness of an initial current sheet,
 $\delta_y$, is set to be $1.2\ \lambda_{\rm i}$.
 The density is given by
 \begin{equation}
  n(y) = n_0 + n_c\cosh^{-2}\left(\frac{y}{\delta_y}\right)
   + n_c\cosh^{-2}\left(\frac{y-y_c'}{\delta_y}\right).
 \end{equation}
 The ratio of the density at the center of the initial current sheet
 to the background one, i.e., $n_c/n_0$, is set to be $4$.
 We assume the uniform electron temperature,
 $T_e=5\times 10^{-3}$
 ($\beta_{e,0}=10^{-2}$ in the initial lobes).
 Ions consist of two components: The current sheet ions and
 the background ions.
 The current sheet ion temperature
 is set to be $T_{i,c}=0.12$, which satisfies the
 relation, $n_c (T_{i,c}+T_{e})=B_0^2/(8\pi)$.
 As for the background ion temperature, we
 assume $T_{i,0}=5\times 10^{-4}$
 ($\beta_{i,0}=10^{-3}$ in the initial lobes).
 Both the current sheet and the background ion temperatures
 are given isotropically.
 We initially input $160$ super-particles per cell at the center of the
 current sheet (i.e., $32$ super-particles per cell at two lobes).

 \section{Results}

 First we shall show the whole structure of reconnection, focusing on
 discontinuities formed along reconnection layers.
 Next, the nature of such discontinuities in kinetic plasmas are
 investigated in accordance with MHD Rankine-Hugoniot relations in
 anisotropic plasmas.

    \paragraph{Structure of reconnection layer}

    In Figure 1, we show the whole structure of reconnection at time
    $t=415\ \Omega_{\rm i}^{-1}$.
    Magnetic field lines continuously reconnect with each other at the
    center of the simulation box and two pairs of reconnection layers are
    formed.
    In the region $\left|x\right|< 70\ \lambda_{\rm i}$,
    ions experience Speiser-type trajectories \citep{Speiser1965,Nakamura1998} and form
    a thin current sheet whose half thickness is about an ion inertial
    length.
    In
    $70\ \lambda_{\rm i}<\left|x\right|< 110\ \lambda_{\rm i}$,
    the current structure becomes in some degree turbulent and its half
    thickness reaches about $5\ \lambda_{\rm i}$.
    Then, the current appears to be concentrated in the areas along
    the PSBL as is predicted by MHD simulations.

    Figure 2 shows enlarged views of the reconnection layers at $x<0$.
    From the top to the bottom, magnetic field lines, 
    the out of plane magnetic field $B_z$,
    flow vectors, the mean ion temperature
    $\left<T_{\rm i}\right> = (T_{{\rm i},\parallel} + 2T_{{\rm i},\perp})/3$,
    and the ion temperature ratio
    $T_{{\rm i},\parallel}/T_{{\rm i},\perp}$
    are shown.
    Vertical red dashed lines separate two regions: Region 1 and
    Region 2 as indicated in Figure 2(b).
    In Region 1, the reconnection layer reaches almost steady state and its
    global structure does not change over time except for small scale turbulent structures.
    In Region 2, the reconnection jet encounters the preceding plasmoid and
    the plasma flows diverge.
    At this time, the region
    $\left|x\right|< 150\ \lambda_{\rm i}$
    is filled with plasmas which originate from two lobes.

    As is seen in Figure 2(a), magnetic field lines begin to bend from
    $x\sim -115\ \lambda_{\rm i}$
    and pile up from $x\sim -125\ \lambda_{\rm i}$.
    The out of plane magnetic field, $B_z$, shows a clear quadrupole
    signature at $-50\ \lambda_{\rm i}<x$ \citep{Hesse1994, Nakamura1998}.
    In $-120\ \lambda_{\rm i}<x<-50\ \lambda_{\rm i}$, the oscillations of
    $B_z$ appear near the central plasma sheet (CPS) \citep{Karimabadi1999}.
    The outflow velocity at $y=0$ is $\sim 0.7$ $V_{A0}$ in Region 1
    and $\sim 0.5$ $V_{A0}$ in Region 2.
    In the piled-up region ($x<-125\ \lambda_{\rm i}$), ions are heated and the ion
    temperature anisotropy considerably decreases as is seen in Figure 2(d)
    and 2(e),
    while in $-120\ \lambda_{\rm i}< x$ (Region 1) the ion temperature
    anisotropy remains high especially in the transition region (the region
    between upstream and downstream).
    The ratio $T_{{\rm i},\parallel}/T_{{\rm i},\perp}$
    is about 4--6 and in good agreement with \cite{Lin1996} and
    \cite{Lottermoser1998}.

    Figures 3(a)--(c) show three different ion velocity distribution
    functions $f(v_x,v_y)$ in the upstream, transition, and
    downstream regions, respectively.
    These ion velocity distribution functions are constructed by using
    super-particles within the white squares shown in Figure 2(e).
    In the upstream region, only cold lobe ions exist with
    the ${\bf E}\times{\bf B}$ drift velocity
    $V_y\simeq cE_zB_x/B^2\simeq -0.15\ V_{A0}$.
    In the transition region, both cold lobe ions and PSBL beam ions exist.
    Characteristic speed of the PSBL beam ions is
    $V_{\rm ibeam}\sim 1.2\ V_{A0}$, and it is known that these PSBL
    ions are accelerated in and around the diffusion region
    \citep{Hoshino1998}.
    Here, note that the ion temperature evaluated from these two components
    in Figure 3(b) gives $T_{{\rm i},\parallel}>T_{{\rm i},\perp}$ in the
    transition region.
    In the downstream region, ions are considerably heated, and the
    shifted-Maxwellian distribution with the outflow velocity
    $V_x\simeq -0.7\ V_{A0}$ is observed.

    Next, to investigate fine structures along reconnection layers
    in more detail, Figures 4--6 respectively show the cross-sectional views
    of discontinuities at $x = -88.3\ \lambda_{\rm i}$, $-115.0\ \lambda_{\rm i}$
    and $-145.0\ \lambda_{\rm i}$ of Figure 2.
    (These locations are indicated by black arrows at the bottom of Figure 2(e).)
    The horizontal axes correspond to $y$ axes of Figure 2, which are normal
    to the initial current layer.
    From the left top to the right bottom, the ion density, bulk velocities
    $V_x$ and $V_y$, the total magnetic field strength, $x$ and $z$
    components of the magnetic field, the current along the initial current,
    the ion temperature ratio
    $T_{{\rm i},\parallel}/T_{{\rm i},\perp}$, and the anisotropic
    parameter $\epsilon\equiv 1 - (\beta_{\parallel}-\beta_{\perp})/2$ are
    shown, respectively.
    The anisotropic parameter $\epsilon$ is useful to
    discuss the net effects of the temperature anisotropy.
    The vertical dash-dotted line stands for the boundary between the
    upstream and transition regions.
    The vertical dotted line stands for the
    boundary between the transition and downstream regions.
    The main judgmental standard points to determine these boundaries are the
    changes of both outflow and inflow bulk velocities.
    Note that the horizontal dashed lines shown in both
    $T_{{\rm i},\parallel}/T_{{\rm i},\perp}$- and $\epsilon$-plots stand for
    the isotropic temperature baselines.

    At $x = -88.3\ \lambda_{\rm i}$, the current $J_z$ concentrates in the
    CPS, even though slow shock-like discontinuities are formed along the
    reconnection layers.
    At $x = -115\ \lambda_{\rm i}$, the enhancement of $J_z$ is seen around
    the transition region.
    Such bifurcated structures of the current indicate that most of
    the magnetic field energy is converted into kinetic and thermal energy of
    plasmas not in the CPS but near the transition regions.
    As for the magnetic field, the out of plane magnetic field $B_z$ is
    confined to the transition region.
    The changes of the density, the bulk velocity, and the magnetic field
    also show the slow shocks-like behavior across these discontinuities.
    At $x = -145\ \lambda_{\rm i}$, a pair of current layers
    can be seen clearly.
    Note that the downstream of the pair of the current layers is the
    plasmoid, and that the inherent heating mechanism would be
    different from that in the quasi-steady state region (Region 1).
    In Region 2, since the radius of curvature of the magnetic field at the CPS
    becomes as large as the ion gyro-radius, the ion motion becomes
    stochastic and would contribute to the plasma heating
    \citep{Buchner1989,Lottermoser1998}.

    As is shown in Figures 4--6, $T_{{\rm i},\parallel}$ is larger than
    $T_{{\rm i},\perp}$ in the transition region.
    The ion temperature anisotropies are reduced across the transition
    region, but we can observe a finite ion temperature anisotropy in the
    downstream region for $x=-88.3\ \lambda_{\rm i}$.
    In the downstream region, since plasma betas are
    larger than those in the transition region, net anisotropic effects are
    significant and may affect the structure of discontinuities and the plasma
    instability.
    Also it should be noted from $\epsilon$-plots in Figures 4--6 that 
    this net anisotropic effect of ions becomes large with decreasing the
    distance from the neutral point.
    In $-100\ \lambda_{\rm i}<x$, the downstream temperature anisotropy
    is large and the fire-hose unstable condition is often satisfied as is
    shown in the $\epsilon$-plot of Figure 4.
    (Note that the fire-hose unstable condition is
    $p_\parallel-p_\perp>B^2/(4\pi)$, i.e.,
    $\epsilon<0$.)
    At $x=-115\ \lambda_{\rm i}$ (Region 1) and
    at $x=-145\ \lambda_{\rm i}$ (Region 2), the ion temperatures are
    almost isotropic.
    From the point of view of RH relations in anisotropic plasmas, such
    downstream temperature anisotropy is important to discuss the
    discontinuities \citep{Lyu1986,Karimabadi1995A}.
    This effect in magnetic reconnection is discussed in the following
    subsection in more detail.

    \paragraph{Comparison with RH relation in anisotropic plasmas}

    We shall investigate the nature of these discontinuities formed
    along reconnection layers by using RH relations in anisotropic plasmas
    \citep{Karimabadi1995A}.
    The basic equations are as follows:
    \begin{eqnarray}
     \left[
      \rho V_n
	    \right]^1_2
     & = 0, \\
     \left[
      \rho V_n^2 + \bar{p} + 
      \frac{1}{3}\left(\epsilon + \frac{1}{2}\right)\frac{\left|{\bf B}\right|^2}{4\pi}
      -\epsilon\frac{B_n^2}{4\pi}
		       \right]^1_2
     & = 0, \\
     \left[
      \rho V_n V_t - \epsilon\frac{B_n B_t}{4\pi}
	    \right]^1_2
     & = 0, \\
     \left[
      \rho V_n\left(V^2 + \frac{\gamma}{\gamma-1}\frac{\bar{p}}{\rho}\right)
      +\frac{\epsilon + 2}{3}V_n\frac{\left|{\bf B}\right|^2}{4\pi}
      -\epsilon V_n\frac{B_n^2}{4\pi}
      -\epsilon V_t\frac{B_nB_t}{4\pi}
	    \right]^1_2
     & = 0,
    \end{eqnarray}
    where the brackets represent $\left[A\right]^1_2=A_1-A_2$ and the
    subscripts, $1$ and $2$, represent upstream and downstream,
    respectively.
    $A_n$ and $A_t$ respectively stand for the normal and tangential
    components of $A$.
    In above equations, the total pressure
    $\bar{p}=\left(p_\parallel+2p_\perp\right)/3$ is introduced instead
    of using the double-adiabatic theory.
    (In more detail, see \cite{Karimabadi1995A}.)

    It is known that the modified intermediate Mach number
    $M_n\equiv V_n/(\sqrt{\epsilon}V_{An})$ is useful to discuss the above
    RH solution, where $V_{An}=B_n/\sqrt{4\pi\rho}$ is the
    Alfv$\acute{\rm e}$n velocity normal to the shock front \citep{Hau1989}.
    Then, the relation between the upstream modified intermediate Mach
    number $M_{n1}$ and its downstream value $M_{n2}$ are obtained, after
    some algebraic calculations, as follows:
    \begin{eqnarray}
     \Lambda_a(\epsilon_2,\theta_1,M_{n2}^2)\cdot\epsilon_1^2M_{n1}^4
      + 2\Lambda_b(\epsilon_1,\epsilon_2,\theta_1,\beta_1,M_{n2}^2)
      \cdot\epsilon_1M_{n1}^2
      + \Lambda_c(\epsilon_1,\epsilon_2,\theta_1,\beta_1,M_{n2}^2)
      & = 0,
    \end{eqnarray}
    where
    \begin{eqnarray*}
     \Lambda_a & = &
      \frac{\gamma-1}{\gamma}\cdot\frac{\xi_2}{\cos^2\theta_{1}}
      - \xi_1M_{n2}^2\tan^2\theta_{1}, \\
     \Lambda_b & = &
     \xi_2\left[
	   \frac{\gamma-1}{\gamma}\cdot\frac{2\left(1-\epsilon_1\right)}{3\cos^2\theta_1}
	   + \frac{\epsilon_1\beta_1}{2\cos^2\theta_1}-\epsilon_2M_{n2}^2
	     \right]
     + \epsilon_1\xi_1M_{n2}^2\tan^2\theta_{1}, \\
     \Lambda_c & = &
     M_{n2}^2\left\{
	      \epsilon_2^2\xi_2\left[
				\frac{\gamma+1}{\gamma}M_{n2}^2
				- \frac{\epsilon_1\beta_1}{\epsilon_2\cos^2\theta_1}
				+ \left(\frac{\epsilon_1}{\epsilon_2}-1\right)
				  \right.
		   \right. \\
     & & \left.
	  \left.
	   + \frac{2}{3}\left(1-\frac{1}{\epsilon_2}\right)
	   \left(\frac{2\gamma-2}{\gamma}-\tan^2\theta_1\right)
	     \right]
	  - \epsilon_1^2\xi_1\tan^2\theta_1
	      \right\},
    \end{eqnarray*}
    and
    \begin{eqnarray*}
     \xi_1 & = &
      \frac{\gamma-1}{\gamma}\left(M_{n2}^2-2 + \frac{1}{\epsilon_2}\right)
      - \frac{1}{3\gamma}\left(2 + \frac{1}{\epsilon_2}\right), \\
     \xi_2 & = &
     \left(M_{n2}^2-1\right)^2.
    \end{eqnarray*}
    $\theta_1$ is the angle between the shock normal and the upstream
    magnetic field line.
    So four parameters, i.e., anisotropic parameters in both upstream and
    downstream regions, an upstream shock angle, and an upstream plasma beta
    are necessary to determine above RH relations.
    These parameters are calculated from physical quantities obtained by our
    simulation and we can obtain the relation between the upstream and downstream
    Alfv$\acute{\rm e}$n Mach numbers.
    Then, if one chooses a certain upstream Alfv$\acute{\rm e}$n Mach
    number, one obtains corresponding downstream Alfv$\acute{\rm e}$n Mach
    numbers.
    It should be noted that since plasma betas in two lobes are much smaller
    than unity, the upstream anisotropic parameter $\epsilon_1$ is almost
    equivalent to unity as is shown in Figures 4--6.
    
    In Figure 7 we show such parameters in magnetic reconnection.
    From the top to the bottom, boundaries separating the
    upstream, transition, and downstream regions in the $x$-$y$ plane, the
    inflow velocity, the upstream shock angle, and the downstream
    anisotropic parameter are shown, respectively.
    To eliminate the effects of initial current plasmas, we do not
    analyze the region $x< -150\ \lambda_{\rm i}$.
    In Figure 7(a), we choose error bars for boundaries so that changes of both 
    $V_x$ and $V_y$ are within them.
    Error bars for $\theta_1$, $V_{n1}$, and $\epsilon_2$ stand for standard
    variations due to the averaging procedure for both upstream and downstream
    regions.
    In order to determine the upstream shock angle, we first evaluate the angle
    between the CPS, i.e., $y=0$, and a shock surface, $\theta_{SC}$, by the
    method of the least square fit.
    Then, using magnetic field data obtained by our simulation results, we
    calculate the angle between the CPS and the upstream magnetic field.
    Finally, we obtain the upstream shock angle $\theta_{1}$, and the
    upstream velocity normal to the shock front, i.e.,
    $\left|V_{n1}\right|=\left|V_x\right|\sin\theta_{SC}+\left|V_y\right|\cos\theta_{SC}$.
    As for the downstream anisotropic parameter, $\epsilon_2$, we eliminated
    points whose standard variations are greater than $0.5$, because of
    difficulties in identifying their
    shock downstream structures.
    
    Since $\theta_{1}$ varies over the range
    of
    $\theta_{1}\simeq 78{}^\circ$--$89{}^\circ$ and the average of inflow
    velocity $\bar{V}_{n1}$ is nearly equal to $0.12$ from Figures 7(b) and
    7(c), we can suppose that the upstream Alfv$\acute{\rm e}$n Mach number
    normal to the shock surface,
    $M_{n1}\equiv V_n/\left(\sqrt{\epsilon_1}V_A\cos\theta_{Bn}\right)$, varies in
    the range $M_{n1}\geq 0.58$ ($M_{n1}^2\geq 0.33$).
    As for the downstream anisotropic parameter, $\epsilon_2$ increases with
    increasing the distance from the magnetic neutral point
    in Region 1, and $\epsilon_2$ is nearly equal to
    unity everywhere in Region 2.

    Now, we know $\epsilon_2$, $\theta_1$, and $M_{n1}$.
    In addition, the upstream plasma betas are nearly equal to $10^{-2}$
    everywhere in the upstream region.
    Therefore, we can draw RH shock solution curves.
    Figure 8 shows such RH solutions in three cases:
    $(\beta_1,\theta_1,\epsilon_2)=(10^{-2},84{}^\circ,0.60)$,
    $(10^{-2},84{}^\circ,0.85)$, and $(10^{-2},84{}^\circ,1.00)$.
    Since there is no recognizable dependence of $\theta_1$ on the nature of
    shock solution curves, $\theta_1$ is supposed to be $84{}^\circ$ here.
    The curves in the region where $M_{n1}\leq 1.0$ correspond to slow
    shock solutions.
    As is shown in the earlier works \citep{Lyu1986,Karimabadi1995A}, slow
    shock solutions are sensitive to the temperature anisotropy and the
    solution curves stretch as $\epsilon_2$ becomes large.
    From this figure, one can find that if $\epsilon_2$ is larger than
    $0.85$, the minimum $M_{n1}^2$ value
    obtained by our simulation, i.e., $M_{n1}^2=0.33$ ($M_{n1}^2\geq 0.33$),
    always has intersection points with the RH shock soolution curve and slow shock
    solutions can exist over the range $0.33\leq M_{n1}^2\leq 1.0$.
    On the other hand in the case of $\epsilon_2<0.85$, the area, where slow shock
    solutions can exist, shrinks and it becomes harder for slow shocks to
    have their solutions with given Mach numbers.

    Based on the above discussion, let us examine Figure 7(d)
    again.
    The horizontal dash-dotted line corresponds to $\epsilon_2=0.85$.
    In the region $-115\ \lambda_{\rm i}< x$, since $\epsilon_2$ is
    smaller than $0.85$, the range of upstream Alfv$\acute{\rm e}$n Mach
    numbers where slow shocks can exist becomes narrower.
    While in the outer region $x\leq -115\ \lambda_{\rm i}$, the ion
    temperature anisotropy is small and $\epsilon_2$ is nearly equal
    to unity.
    It enables slow shocks to exist stationarily in the calculated upstream
    Alfv$\acute{\rm e}$n Mach numbers.
    The compression ratio $\rho_2/\rho_1$ in such regions
    [simulation/theory (error \%)]
    can be calculated as
    [1.9/2.5 (32 \%)] at $x\sim -115\ \lambda_{\rm i}$ (Region 1)
    with $M_{n1}=0.95$, $\theta_1=82.8^\circ$, $\beta_1=10^{-2}$, and
    $\epsilon_2=0.98$,
    and
    [2.1/2.6 (24 \%)] at $x\sim -145\ \lambda_{\rm i}$ (Region 2)
    with $M_{n1}=1.0$, $\theta_1=87.5^\circ$, $\beta_1=10^{-2}$, and
    $\epsilon_2=0.95$.

 \section{Discussion and Conclusion}

 We have discussed the relation between the ion temperature anisotropy
 obtained by our two-dimensional kinetic simulation of magnetic
 reconnection and the formation of slow shocks in accordance with the
 RH relations.
 From the point of view of RH relations in anisotropic plasmas, the
 parameters to determine shock solutions are the upstream plasma beta
 $\beta_1$ and shock angle $\theta_1$, the downstream ion temperature
 anisotropy $\epsilon_2$, and the upstream Alfv$\acute{\rm e}$n Mach number
 $M_{n1}$.
 Among these four parameters, $\beta_1$, $\theta_1$, and $\epsilon_2$
 have particularly an important influence on the nature of RH shock
 solution curves.
 We have evaluated the spatial profiles of these parameters in magnetic
 reconnection and discussed whether or not the discontinuities satisfy
 the conditions for slow shocks.
 In this study, it has been shown that the downstream ion temperature
 anisotropy along the reconnection layer decreases
 with increasing distance
 from the neutral point, and that a pair of current layers
 is formed in the region where plasmas are considerably isotropized
 ($\left|x\right|\geq 115\ \lambda_{\rm i}$).
 This spatial distribution of the downstream ion temperature anisotropy
 is strongly related to the formation of slow shocks, and the
 relaxation of the ion temperature anisotropy allows RH shock solutions
 in a broad range of upstream Alfv$\acute{\rm e}$n Mach numbers in
 collisionless magnetic reconnection.
 Let us discuss the dependence of such parameters on the formation of
 slow shocks in magnetic reconnection in more detail.

 First, we will refer to effects of downstream temperature anisotropy.
 Previous studies for the RH solutions in anisotropic plasmas
 suggest that the most important parameter for the RH relations is the
 downstream anisotropic parameter $\epsilon_2$ \citep{Lyu1986,Karimabadi1995A}.
 In collisionless magnetic reconnection, such temperature anisotropy is
 due to the PSBL beam ions, whose characteristic bulk velocity is about the
 lobe Alfv$\acute{\rm e}$n velocity.
 Plasma mixing between cold lobe plasma,
 whose bulk velocity is about a tenth of the lobe
 Alfv$\acute{\rm e}$n velocity, and PSBL ion beam components result
 in high temperature parallel to the magnetic field.
 Under these circumstances, the downstream anisotropic parameter
 $\epsilon_2$ becomes smaller than unity.
 Here, we would emphasize that the generation mechanism of the ion temperature
 anisotropy in magnetic reconnection is quite different from that in the
 slow shocks without magnetic reconnection.
 In case of slow shocks without reconnection, 
 incident ions and backstreaming ions from the shock
 downstream region are known to form the ion temperature anisotropy.
 On the other hand, in the case of slow shocks with magnetic reconnection, 
 in addition to the backstreaming ions, PSBL ion beams accelerated around the
 diffusion region can contribute to the ion temperature anisotropy as well
 \citep{Hoshino1998}.
 Getting back to the diagram of $M_{n2}^2$-$M_{n1}^2$ plot discussed before,
 as $\epsilon_2$ becomes small, the region where slow shock
 solutions exist shrinks.
 This is the main reason why slow shocks are hard to have their solutions
 near the diffusion region in magnetic reconnection.
 
 In addition to the formation of such temperature anisotropy, a
 particular interest is why the relaxation of the ion temperature
 anisotropy occurs away from the diffusion region.
 In collisionless magnetic reconnection, the anisotropic PSBL ions
 can become thermalized by the
 Alfv$\acute{\rm e}$nic waves generated by the ion-cyclotron beam instability,
 the ion/ion cyclotron instability \citep{Winske1984},
 and the EMIIC (electromagnetic ion/ion cyclotron) instability
 \citep{Omidi1990}.
 From observations in the Earth's magnetotail, the existence of the
 ion cyclotron beam instability in the PSBL has been confirmed
 \citep{Kawano1994,Takada2005}.
 Additionally, the fire-hose instability can take place in the
 reconnection exhaust \citep{Karimabadi1999,Liu2011}.
 Our simulation results also support the fire-hose instability enhanced
 near the CPS.
 As shown in Figures 4--6, since the temperature
 anisotropic parameter $\epsilon$ is greater than $0$, the
 fire-hose instability is absent in the transition region.
 However, near the CPS, the condition $\epsilon<0$ for the fire-hose instability can be
 satisfied in $-100\ \lambda_{\rm i}< x$.
 We think that these instabilities may play an important role in the
 relaxation process of the downstream ion temperature anisotropy.
 As a result, Maxwellian-like distribution functions of ions are observed away
 from the neutral point.
 From the viewpoint of anisotropic RH relations, such relaxation of the
 downstream temperature anisotropy away from the diffusion region enables slow
 shocks to have their solutions in wider range of the upstream
 Alfv$\acute{\rm e}$n Mach numbers.

 We briefly refer to the upstream shock angle, $\theta_1$.
 From our simulation results, $\theta_1$ is more or less in the range
 of
 $78{}^\circ$--$89{}^\circ$.
 According to the anisotropic RH relations, such a
 variation obtained by our simulation does not result in a recognizable
 impact on the nature of solution curves.
 However, if $\theta_1$ is smaller than $78{}^\circ$ by keeping the
 inflow velocity of $0.12$, the upstream 
 Alfv$\acute{\rm e}$n Mach number becomes in the range of $M_{n1}<0.58$
 ($M_{n1}^2< 0.33$) and the slow shock condition is not satisfied.

 In this paper, we have shown the existence of slow shock
 discontinuities and the resultant current sheet profiles
 in both Regions 1 and 2, but the behavior of the downstream
 temperature anisotropies are different between two regions.
 The downstream ion temperature in Region 2 is nearly isotropic, while
 that in Region 1 is rather high even though the temperature anisotropy
 can be relaxed
 with increasing distance
 from the X-type neutral point.
 This behavior seems to be important for the satellite observation
 of the slow mode shock.  So far the observational study of the 
 slow shock detection in the Earth's magnetotail assumed the isotropic
 temperature \citep{Saito1995,Seon1995}, but the 
 RH study of the slow mode shock including the temperature anisotropy 
 may distinguish the slow shock region between Regions 1 and 2.

 In Region 1,
 we found the relaxation of the anisotropic temperature and the 
 formation of the slow shock, but 
 one might indicate that larger scale simulations would
 result in more isotropic plasma distribution in the slow shock region.
 In fact, we studied larger scale simulations,
 but we found that as time goes on, 
 other
 magnetic islands are formed from the diffusion region, 
 grow, and are ejected into the outflow region.
 As a result, the size of Region 1 cannot become
 larger than $\sim 120\ \lambda_{\rm i}$.

 We obtained that the minimum distance for relaxation of the temperature anisotropy
 required for the formation of slow shock is about $115$ times ion inertia 
 length based on our two-dimensional hybrid simulation.  However, 
 some other processes that are not included in our simulation may quickly
 reduce the ion temperature anisotropy in a shorter spatial scale from the 
 X-type neutral point.   
 Such candidates might be the three
 dimensionality (e.g., drift-mode instabilities), or the instabilities
 due to electron kinetic effects which cannot be considered in our
 hybrid model.
 Recently \cite{Yin2007} performed the oblique slow shocks by
 full-particle simulations, and discussed how electron kinetic effects are
 related to kinetic Alfv$\acute{\rm e}$n waves and could alter the
 structure of slow shocks.
 These possibilities will be investigated in future works.


%
%
%
%
%
%

%
%
%
%

\begin{acknowledgments}
 This work was supported by the editing assistance from the GCOE
 program.
 We thank Masaki Fujimoto, Lin-Ni Hau, Mariko Hirai, and Ryo Yoshitake
 for useful discussions.
 We would like to express our gratitude to the two referees for a number
 of useful comments to improve this paper.
\end{acknowledgments}

%
%
%
%
%
%
%
%
%
%





%
%

   \begin{figure*} 
    \noindent\includegraphics[width=39pc]{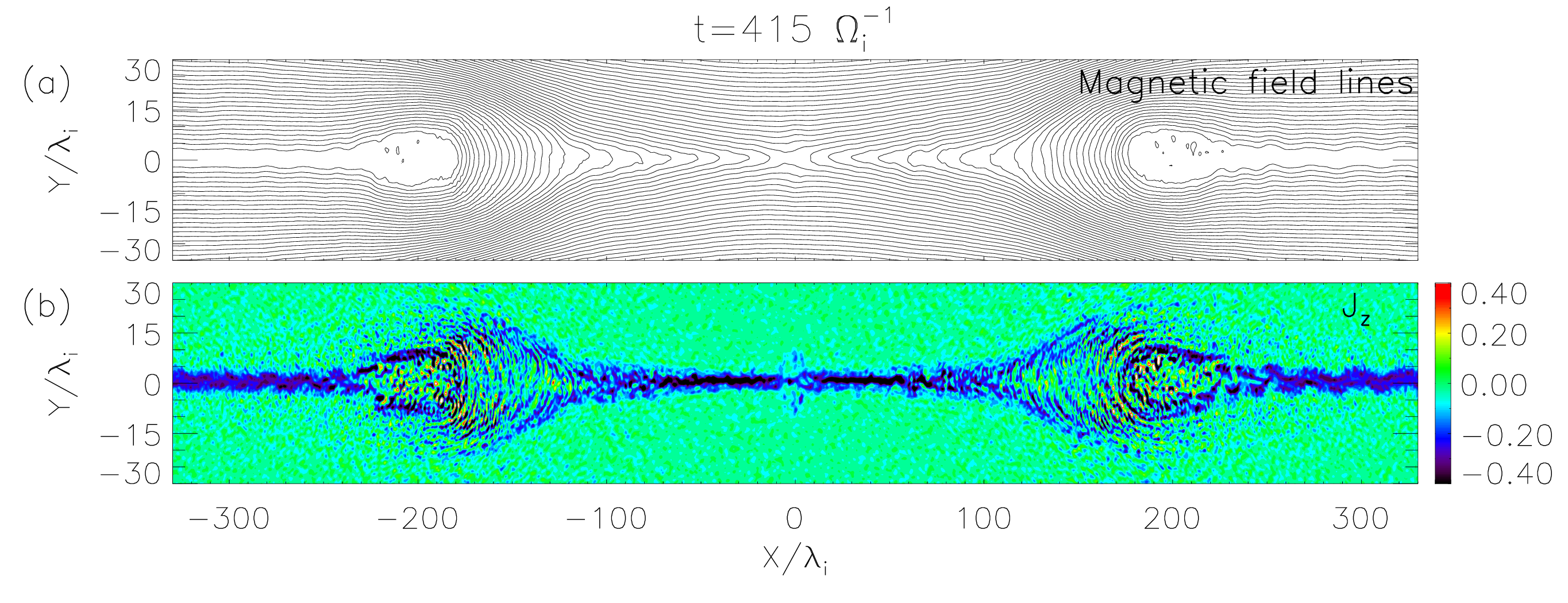}\\
    \caption{
    (a) Magnetic field lines and (b) the electric current in the
    $z$ direction at time $t=415\ \Omega_{\rm i}^{-1}$.
    The initial current is in the $-z$ direction.
    }
   \end{figure*}
   \begin{figure*} 
    \noindent\includegraphics[width=32pc]{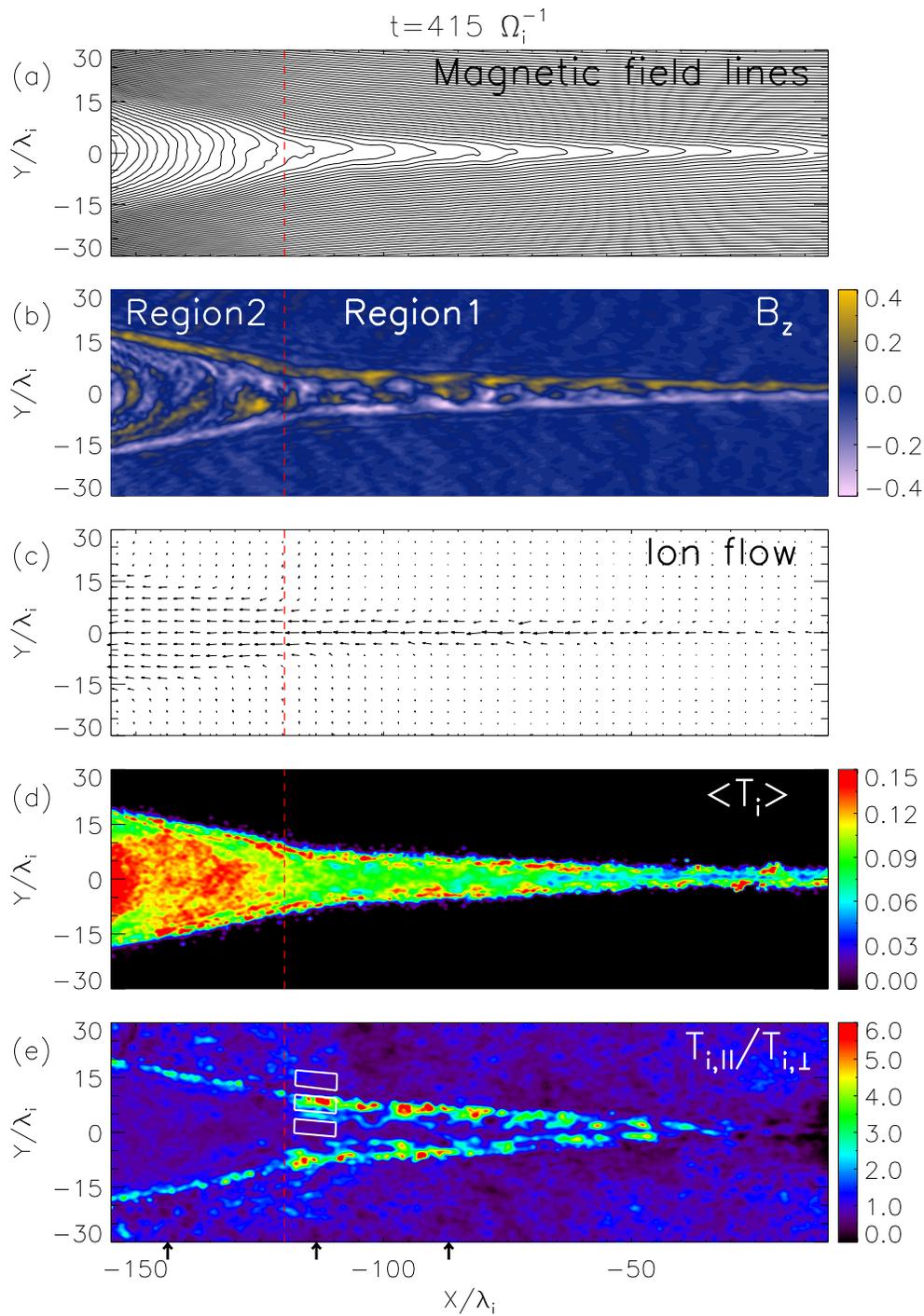}\\
    \caption{
    Enlarged views of the reconnection layer at $x<0$.
    (a) Magnetic field lines, (b) the out of plane magnetic field
    $B_z$, (c) the ion flow vector, (d)
    the ion temperature, and (e) the ion temperature ratio
    $T_{{\rm i},\parallel}/T_{{\rm i},\perp}$ are shown in the $x$-$y$
    plane.
    }
   \end{figure*}
   \begin{figure*} 
    \noindent\includegraphics[width=35pc]{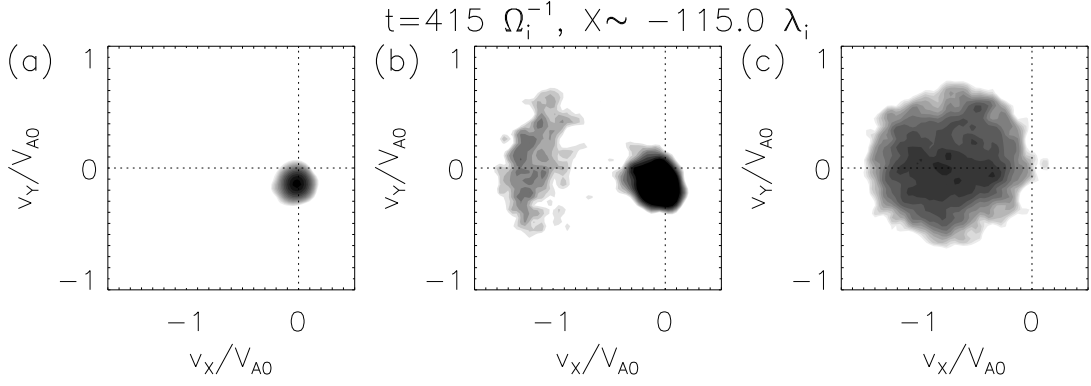}\\
    \caption{
    Ion velocity distribution
    functions ($f(v_x,v_y)$) in the upstream, transition, and downstream
    regions. Their locations are indicated as white squares in Figure 2(e).
    }
   \end{figure*}
   \begin{figure*} 
    \noindent\includegraphics[width=35pc]{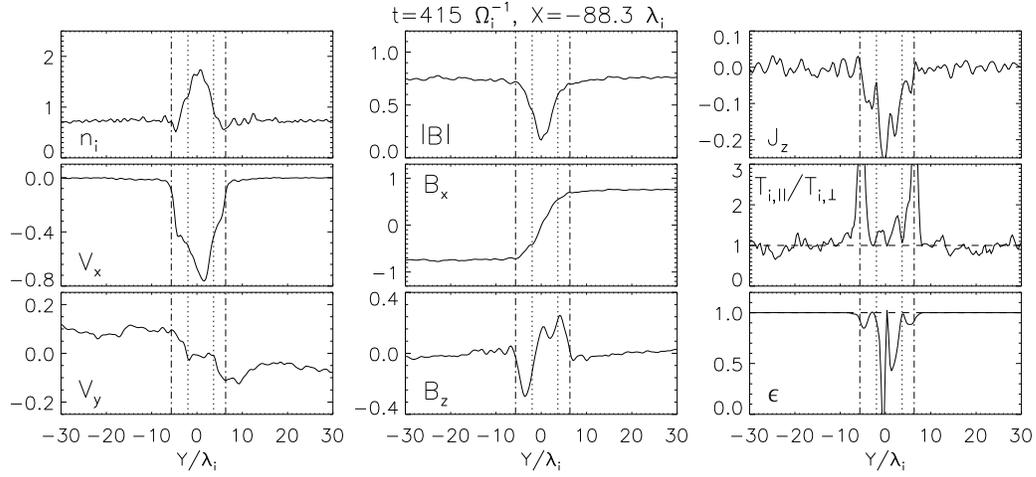}\\
    \caption{
    Cross-sectional views of discontinuities at
    $x = -88.3\ \lambda_{\rm i}$.
    }
   \end{figure*}
   \begin{figure*} 
    \noindent\includegraphics[width=35pc]{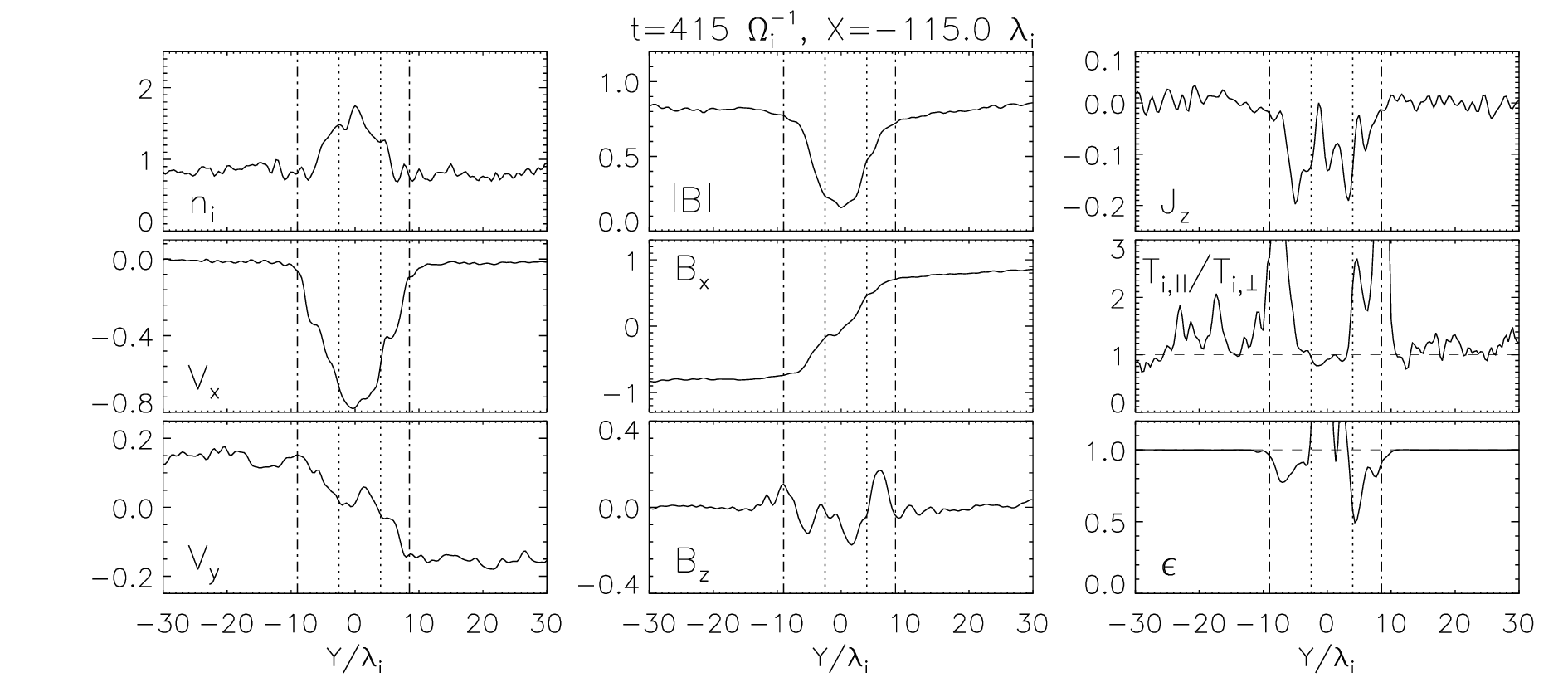}\\
    \caption{
    Cross-sectional views of discontinuities at
    $x = -115.0\ \lambda_{\rm i}$.
    }
    \end{figure*}
   \begin{figure*} 
    \noindent\includegraphics[width=35pc]{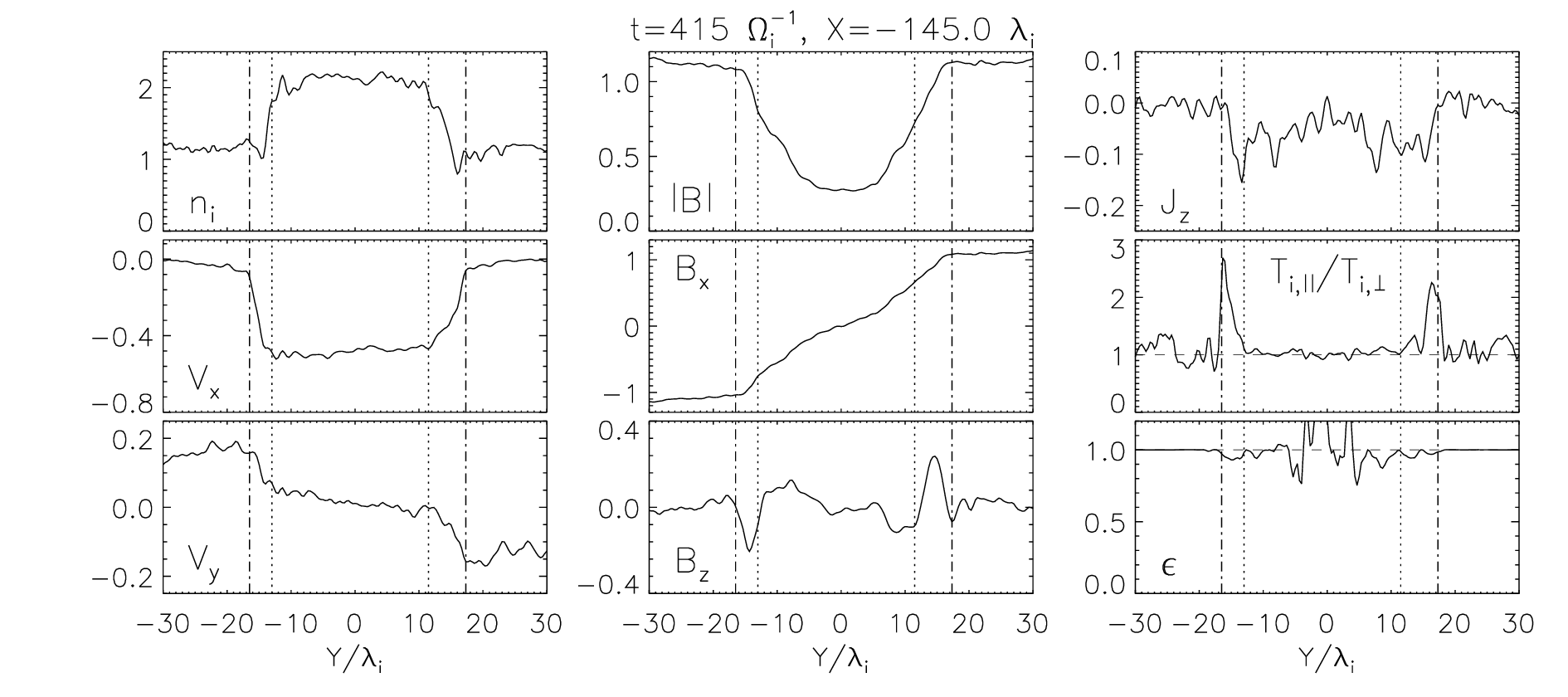}\\
    \caption{
    Cross-sectional views of discontinuities at
    $x = -145.0\ \lambda_{\rm i}$.
    }
   \end{figure*}
   \begin{figure*} 
    \noindent\includegraphics[width=35pc]{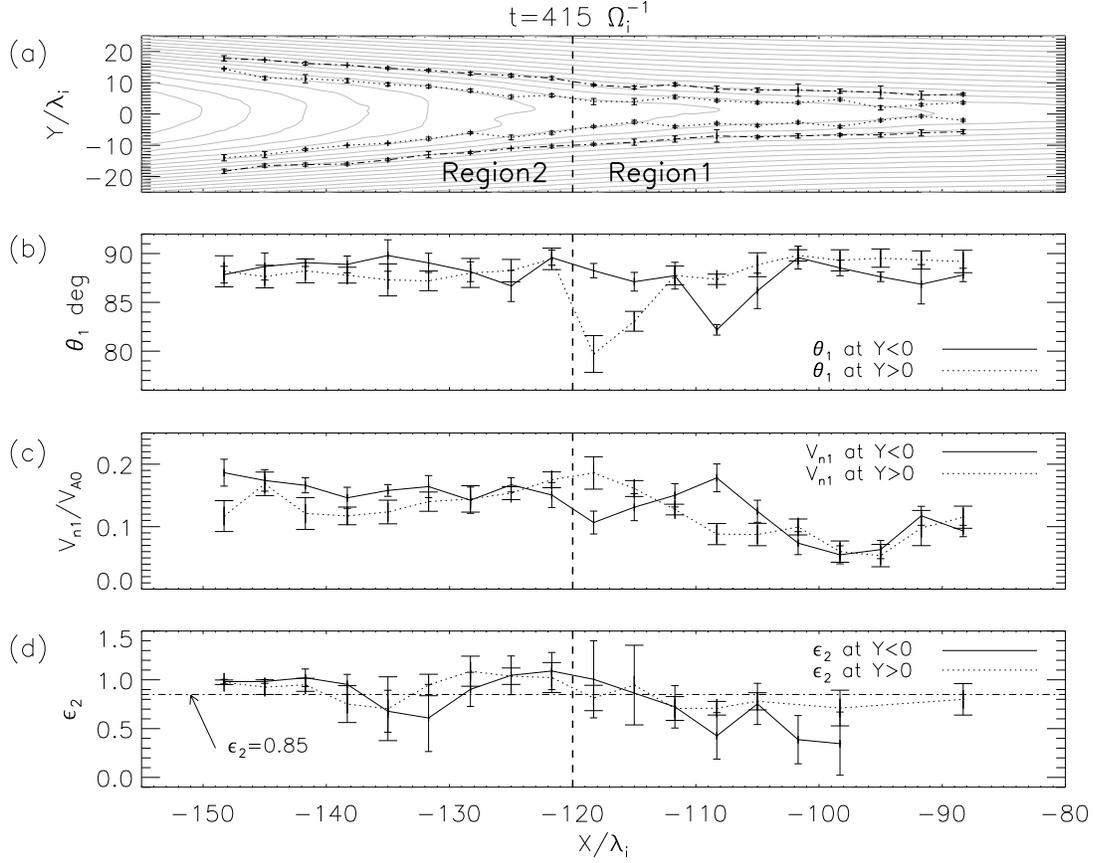}\\
    \caption{
   (a) Dash-dotted line and dotted line respectively stand for
   the boundaries separating the upstream and transient regions, and the
   ones separating the transient and downstream regions. In addition,
   spatial profiles of (b) the upstream shock angle, (c) the upstream
   velocity normal to the surfaces of discontinuities, and (d) the
   downstream anisotropic parameter are shown as a function of the
   distance from the neutral point.
    }
   \end{figure*}
    \begin{figure*} 
    \noindent\includegraphics[width=20pc]{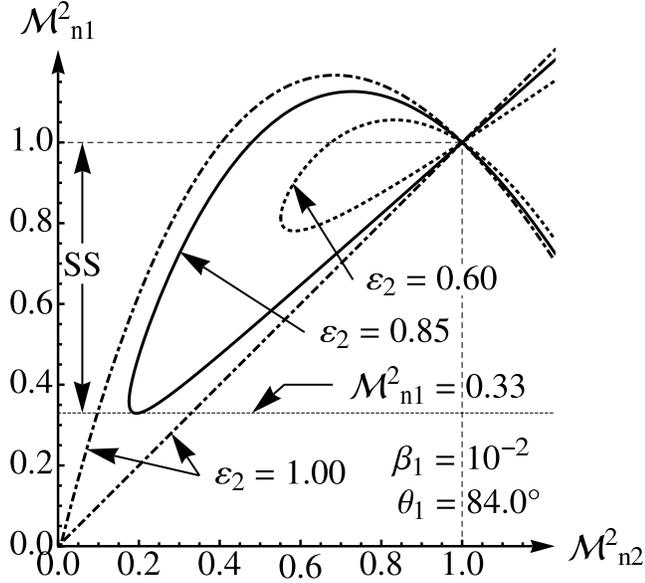}\\
    \caption{
    $M_{n2}^2$-$M_{n1}^2$ plot of RH solutions.
    Isotropic ($\epsilon_2=1$) and anisotropic cases
    ($\epsilon_2=0.85,0.6$) are shown. The upstream plasma beta and the shock
    angle are assumed to be $\beta_1=10^{-2}$ and $\theta_1=84{}^\circ$ in
    all these three cases.
    {\bf The} square of the minimum upstream Alfv$\acute{\rm e}$n Mach number
    calculated in
    our simulation, i.e., $M^2_{n1}=0.33$, {\bf is} shown by the horizontal
    dotted line.
    The area where slow shocks (SS) can exist
    is indicated by a double-headed arrow ($0.33\leq M^2_{n1}\leq 1.0$
    in case of $\epsilon_2\geq 0.85$).
     }
    \end{figure*}
    \pagebreak
\end{article}

\end{document}